\documentclass[12pt]{article}
\usepackage{amssymb}
\def\beq{\begin{equation}}
\def\eeq{\end{equation}}
\usepackage[all,2cell,dvips]{xy}

\def\IR{\relax{\rm I\kern -.18em R}}

\begin{document}
\title{Stability of solitonic solutions of Super KdV equations under Susy breaking conditions }
\author{ \Large  A. Restuccia*, A. Sotomayor**}
\maketitle{\centerline{*Departamento de F\'{\i}sica}}
\maketitle{\centerline{Universidad de Antofagasta }
\maketitle{\centerline{*Departamento de F\'{\i}sica}}
\maketitle{\centerline{Universidad Sim\'on Bol\'{\i}var}

\maketitle{\centerline{**Departamento de Matem\'aticas}
\maketitle{\centerline{Universidad de Antofagasta}}
\maketitle{\centerline{e-mail:arestu@usb.ve, asotomayor@uantof.cl
}}
\begin{abstract} A supersymmetric
breaking procedure for $N=1$ Super Korteweg-de Vries (SKdV),
preserving the positivity of the hamiltonian as well as the
existence of solitonic solutions, is implemented. The resulting
solitonic system is shown to have nice stability properties.
\end{abstract}


Pacs: 12.60.Jv, 02.30.lk, 11.30.-j, 05.45. Yv, 02.30. Jr

\section{Introduction}
The breaking of supersymmetry in physical systems is always an
interesting aspect to analyze. We consider a solitonic system
\cite{Adrian} arising from the breaking of supersymmetry in the
$N=1$ Super KdV system \cite{Mathieu1,Manin}. In the latter there
is only one hamiltonian structure in distinction to the
bi-hamiltonian one in the Korteweg-de Vries (KdV) system. Its
hamiltonian however is not manifestly positive. Nevertheless the
quantum formulation of the theory yields a manifestly positive
definite self-adjoint operator. The stability of the ground state
of the system is then assured from it. We considered in
\cite{Adrian} the supersymmetry breaking of the SKdV system by
changing the Grassmann algebra structure of the SKdV formulation
to a Clifford algebra one. This susy breaking mechanism has
already appeared in several works, see for example
\cite{Seibert}. One may then
 obtain a solitonic system with the same
evolution equation for the new Clifford algebra valued field as
one had for the odd Grassmann algebra valued one in the SKdV
system and what is more important with a bounded from below
hamiltonian. The system presents solitonic solutions although the
infinite set of local conserved quantities of SKdV breaks down. We
will show in this work that this solitonic system has nice
stability properties.

The stability in the sense of Liapunov of the one-soliton
solution of KdV equation was first proven by Benjamin
\cite{Benjamin} and Bona \cite{Bona}. The proof makes use of the
first few conserved quantities of the KdV equation. In particular
the fact that one of then is the square of the $L_2$ norm is
relevant in their argument. The use of conserved quantity was
also considered in a stability argument by Boussinesq
\cite{Boussinesq}. In this paper we make use of this main idea to
prove stability of the one-soliton solution of the coupled
equation, with fields valued on a Clifford algebra, derived from
the supersymmetric breaking of the  $N=1$ SKdV equation.

\section{SKdV and the breaking of supersymmetry}
The fields $u(x,t)$ and $\xi(x,t)$ describing $N=1$ SKdV
equations \cite{Mathieu1} take values on the even and odd part of
a Grassmann algebra respectively. The $N=1$ SKdV equations are

\beq\begin{array}{l}u_t=-u^{\prime\prime\prime}+6uu^\prime-3\xi\xi^{\prime\prime}
\\ \xi_t=-\xi^{\prime\prime\prime}+3{(\xi u)}^\prime.
\end{array}\eeq
This system of partial differential equations has infinite local
conserved charges as well as infinite non-local conserved charges
\cite{Mathieu1,Dargis,Gardner,Andrea1,Andrea2}. The first few
local ones are
\beq\begin{array}{l}H_{\frac{1}{2}}=\int_{-\infty}^{+\infty}\xi
dx,\hspace{3mm}H_1=\int_{-\infty}^{+\infty} u dx
\\ H_3=\int_{-\infty}^{+\infty}\left(u^2-\xi\xi^\prime\right)dx,\hspace{3mm} H_5=\int_{-\infty}^{+\infty}
\left(2u^3+{(u^\prime)}^2-\xi^\prime\xi^{\prime\prime}-4u\xi\xi^\prime\right)dx.
\end{array}\eeq

To break supersymmetry we consider the fields $u$ and $\xi$ taking
values on a Clifford algebra instead of being Grassmann algebra
valued. We thus take $u$ to be a real valued field while $\xi$ to
be an expansion in terms of the generators $e_i,i=1,\ldots$ of
the Clifford algebra: \beq\xi=\sum_i
\varphi_ie_i+\sum_{ij}\varphi_{ij}e_ie_j+\sum_{ijk}\varphi_{ijk}e_ie_je_k+\cdots\eeq
where \beq e_ie_j+e_je_i=-2\delta_{ij}\eeq and
$\varphi_i,\varphi_{ij},\varphi_{ijk},\ldots$ are real valued
functions. We define $ \bar{\xi}=\sum_{i=1}^\infty
\varphi_i\bar{e_i}+\sum_{ij}\varphi_{ij}\bar{e_j}\bar{e_i}+\sum_{ijk}\varphi_{ijk}\bar{e_k}\bar{e_j}\bar{e_i}+\cdots$
where $\bar{e_i}=-e_i$. We denote as in superfield notation the
body of the expansion those terms associated with the identity
generator and the soul the remaining ones. Consequently the body
of $ \xi\bar{\xi}$, denoted by $ \mathcal{P}(\xi\bar{\xi})$, is
equal to
$\Sigma_i\varphi_i^2+\Sigma_{ij}\varphi_{ij}^2+\Sigma_{ijk}\varphi_{ijk}^2+\cdots$
In what follows, without loss of generality, we rewrite
$\mathcal{P}(\xi\bar{\xi})=\Sigma_i
\varphi_i^2+\Sigma_{ij}\varphi_{ij}^2+\Sigma_{ijk}\varphi_{ijk}^2+\cdots$
simply as $\mathcal{P}(\xi\bar{\xi})=\Sigma_i\varphi_i^2$.

The system of partial differential equations arising from the
breaking of supersymmetry which has the required properties as
discussed in \cite{Adrian} is

\beq\begin{array}{l}u_t=-u^{\prime\prime\prime}-uu^\prime-\frac{1}{4}{\left(\mathcal{P}\left(\xi\bar{\xi}\right)\right)}^\prime
\\ \xi_t=-\xi^{\prime\prime\prime}-\frac{1}{2}{(\xi u)}^\prime.
\end{array}\eeq

The system (5) has the following conserved charges

\beq\begin{array}{llll}\hat{H}_{\frac{1}{2}}=\int_{-\infty}^{+\infty}\xi
dx, \\ \hat{H}_1=\int_{-\infty}^{+\infty} u dx,
\\ V\equiv \hat{H}_3=\int_{-\infty}^{+\infty}\left(u^2+\mathcal{P}\left(\xi\bar{\xi}\right)
\right)dx, \\ M\equiv \hat{H}_5=\int_{-\infty}^{+\infty}
\left(-\frac{1}{3}u^3-\frac{1}{2}u\mathcal{P}\left(\xi\bar{\xi}\right)+{(u^\prime)}^2+\mathcal{P}\left(\xi^\prime\bar{\xi}^\prime\right)\right)dx.
\end{array}\eeq
The system has multi-solitonic solutions, for example,
$u(x,t)\equiv \phi(x,t)=3\mathcal{C}\frac{1}{\cosh^2(z)}, z\equiv
\frac{1}{2}\mathcal{C}^{\frac{1}{2}}(x-(1+\mathcal{C})t+a),\xi(x,t)=0,$
where $\phi(x,t)$ is the one-soliton solution of KdV equation, $a$
is an arbitrary real number and $ \mathcal{C}>0$.

This system with a change of sign in the third term of the right
hand member of the first equation of system (5) was considered in
\cite{Olver,Sokolov}.

It will be important in the following stability argument that \beq
\int_{-\infty}^\infty\mathcal{P}\left(\xi\bar{\xi}\right)dx={\|\xi\|}^2_{L_2},\eeq
hence $V$ is the $L_2$ norm of $(u,\xi)$. This property is absent
for the system with a positive sign on the third term of the first
equation in (5).

\section{Stability properties of the system}
We may now analize the stability of the ground state as well as
the stability of the one-soliton solution of the system (5).

We start the analysis by considering an a priori bound for the
solutions of system (5). We denote by $\|\hspace{2mm} \|_{H_1}$
the Sobolev norm:

\beq {\|(u,\xi)
\|_{H_1}}^2=\int_{-\infty}^{+\infty}\left[\left(u^2+\Sigma_{i=0}^\infty
\varphi_i^2 \right)+\left ({u^\prime}^2+\Sigma_{i=0}^\infty
{\varphi_i^\prime}^2\right)\right]dx \eeq

 We obtain \beq {\|(u,\xi)
\|_{H_1}}^2\leq
V+M+\frac{1}{2}\int_{-\infty}^{+\infty}|u|\left(u^2+\mathcal{P}\left(\xi\bar{\xi}\right)\right)dx,
\eeq

we now use

\beq \sup|u|\leq \frac{1}{\sqrt{2}}{\|u \|_{H_1}}\leq
\frac{1}{\sqrt{2}}{\|(u,\xi) \|_{H_1}};\eeq it yields

\beq {\|(u,\xi) \|_{H_1}}^2\leq V+M+\frac{1}{2\sqrt{2}}V{\|(u,\xi)
\|_{H_1}}. \eeq

From (11) it follows

\beq{\|(u,\xi) \|_{H_1}}\leq \frac{d+\sqrt{d^2+4e}}{2}\eeq

where $d=\frac{1}{2\sqrt{2}}V$ and $e=V+M.$ We notice that
$d^2+4e\geq0.$

Consequently, given $V$ and $M$ from the initial data and a
solution  satisfying those initial conditions, the ${\|(u,\xi)
\|_{H_1}} $ is bounded by (12). The a priori bound is a strong
evidence of the existence of the solution for $0\leq t.$ We will
consider this existence problem elsewhere. In this work we assume
the existence of the solution and its continuous dependence on the
initial data under smooth enough assumptions on the initial
perturbation.

We consider the stability in the sense of Liapunov. In particular
we take the same definition as in \cite{Benjamin}: $(
\hat{u},\hat{\xi})$, a solution of (5), is stable if given
$\epsilon$ there exists $\delta$ such that for any solution
$(u,\xi)$ of (5), satisfying at $t=0$

\beq
d_I\left[\left(u,\xi\right),\left(\hat{u},\hat{\xi}\right)\right]<\delta
\eeq

then

\beq
d_{II}\left[\left(u,\xi\right),\left(\hat{u},\hat{\xi}\right)\right]<\epsilon\eeq

for all $t\geq 0$.

$d_I$ and $d_{II}$ denote two distances to be defined.

We consider now the stability problem of the ground state
solution $ \hat{u}=0,\hat{\xi}=0.$

We take $d_I$ and $d_{II}$ to be the Sobolev norm
$\|(u-\hat{u},\xi-\hat{\xi}) \|_{H_1}.$

We get \[V\leq \|(u,\xi) \|^2_{H_1}\] and
\[M\leq \int_{-\infty}^{+\infty}\left(\frac{1}{2}|u|\left(u^2+\mathcal{P}\left(\xi\bar{\xi}\right)\right)+{u^\prime}^2
+\mathcal{P}\left(\xi^\prime\bar{\xi}^\prime\right)\right)dx \leq
\frac{1}{2\sqrt{2}}{\|(u,\xi) \|_{H_1}}^3+{\|(u,\xi)
\|_{H_1}}^2.\]

At $t=0$ we then have, using (13),
\begin{eqnarray*}V&<&\delta^2 \\ M &<& \frac{1}{2\sqrt{2}}\delta^3+\delta^2. \end{eqnarray*}

Consequently, from the a priori bound (12), we obtain for any
$t\geq 0$ \[\|(u,\xi) \|_{H_1}<\epsilon\] for any given
$\epsilon$, provided $\delta$ is conveniently chosen.

This argument proves the stability of the ground state solution.

 We now consider the stability of the
one-soliton solution $ \hat{u}=\phi,\hat{\xi}=0.$ The proof of
stability is based on estimates for the $u$ field which are
analogous to the one presented in \cite{Benjamin,Bona} while a
new argument will be given for the $\xi$ field. The distances we
will use are

\beq
d_I\left[\left(u_1,\xi_1\right),\left(u_2,\xi_2\right)\right]=\|\left(u_1-u_2,\xi_1-\xi_2\right)\|_{H_1}
\eeq \beq
d_{II}\left[\left(u_1,\xi_1\right),\left(u_2,\xi_2\right)\right]=\inf_\tau\|\left(\tau
u_1-u_2,\xi_1-\xi_2\right)\|_{H_1} \eeq where $\tau u_1$ denotes
the group of translations along the $x$-axis. $d_{II}$ is a
distance on a metric space obtained by identifying the
translations of each $u\in H_1( \mathbb{R})$ \cite{Benjamin}.
$d_{II}$ is related to a stability in the sense that a solution
$u$ remains close to $ \hat{u}=\phi$ only in shape but not
necessarily in position.

We first assume that

\beq V(u,\xi)=V( \hat{u},\hat{\xi})=V(\phi,0) \eeq

and

\beq \int_{-\infty}^{+\infty}\xi dx=
\int_{-\infty}^{+\infty}\hat{\xi}dx=0 \eeq

and then we will relax this conditions to get the most general
formulation of the stability problem.

Following \cite{Benjamin} we define

\[h(x,t)=u(x-a,t)-\phi(x+\mathcal{C}t)\] where $a$ is defined, for
each $t\geq0$,by
\[\int_{-\infty}^{+\infty}{\left[u\left(x,t\right)-\phi\left(x+a\right)\right]}^2dx=\inf_{y\in \mathbb{R}}\int_{-\infty}^{+\infty}
{\left[u\left(x,t\right)-\phi\left(x+y\right)\right]}^2dx.\]

In \cite{Bona} it was proven that the infimum is taken on finite
values of $y$.

We obtain,

\beq \begin{array}{l} \Delta M\equiv
M(u,\xi)-M(\phi,0)=\int_{-\infty}^{+\infty}\left[{h^\prime}^2+\left(\mathcal{C}-\phi\right)h^2+\Sigma_{i=0}^\infty{\varphi_i^\prime}^2+
\right.
\\\left. +\left(\mathcal{C}-\frac{1}{2}\phi\right)\Sigma_i\varphi_i^2-\left(\frac{1}{3}h^3-\frac{1}{2}\sum_{i=0}^\infty
\varphi_i^2h\right)\right]dx,
\end{array}\eeq

where we have used $
\mathcal{\mathcal{C}}\left(V(u,\xi)-V(\phi,0)\right)=0$ and that
$\phi$ is the soliton solution of KdV equation and hence it
satisfies \beq
\phi^{\prime\prime}+\frac{1}{2}\phi^2=\mathcal{C}\phi.   \eeq

We denote $\delta^3M\equiv
\int_{-\infty}^{+\infty}\left(-\frac{1}{3}h^3+\frac{1}{2}\sum_{i=0}^\infty
\varphi_i^2h \right)dx.$ We then have

\beq |\delta^3M|\leq \frac{1}{2}\sup
|h|{\|\left(h,\xi\right)\|}_{H_1}^2\leq
\frac{1}{2\sqrt{2}}\|(h,\xi) \|_{H_1}^3 \eeq

where we have used $\sup|h|\leq \frac{\sqrt{2}}{3}\|h\|_{H_1}.$

Coming back to $\Delta M$ we get

\beq \begin{array}{l}|\Delta
M|\leq\int_{-\infty}^{+\infty}\left({h^\prime}^2+\mathcal{C}h^2+\sum_{i=0}^\infty{\varphi_i^\prime}^2+\mathcal{C}\sum_{i=0}^\infty\varphi_i^2\right)dx+
|\delta^3M|\leq \\
\leq\max\left(1,\mathcal{C}\right){\|\left(h,\xi\right)\|}^2_{H_1}+\frac{1}{3\sqrt{2}}{\|\left(h,\xi\right)\|}^3_{H_1}.
\end{array} \eeq

At $t=0$ we will assume that

\beq
d_I\left[\left(u,\xi\right),\left(\phi,0\right)\right]=\|\left(h,\xi\right)\|_{H_1}<\delta
\eeq

hence \beq|\Delta
M|\leq\left[\max\left(1,\mathcal{C}\right)+\frac{1}{3\sqrt{2}}\delta\right]{\|\left(h,\xi\right)\|}_{H_1}^2.\eeq

$\Delta M$ is a conserved quantity on the space of solutions of
the system (5). By taking $\delta$ small enough in (23) we can
make $\Delta M$ as small we wish. The second step in the proof of
stability is to argue that at any $t\geq0$ $\Delta M$ satisfies
the bound \beq \Delta M\geq
Kd_{II}\left[\left(u,\xi\right),\left(\phi,0\right)\right]. \eeq
This bound completes the proof, in fact we can make, at $t=0$,
$|\Delta M|$ as small as we wish and hence we can always satisfy
(14) for any given $\epsilon$.

We decompose $\Delta M$ into \beq \Delta
M=\delta_h^2M+\delta_\xi^2M+\delta^3M \eeq

where
\beq\begin{array}{l}\delta_h^2M=\int_{-\infty}^{+\infty}\left[{h^\prime}^2+\left(\mathcal{C}-\phi\right)h^2\right]dx\\
\delta_\xi^2M=\int_{-\infty}^{+\infty}\left[\sum_{i=0}^\infty{\varphi_i^\prime}^2+\left(\mathcal{C}-\frac{1}{2}\phi\right)
\sum_{i=0}^\infty\varphi_i^2\right]dx.
\end{array} \eeq

Using the same argument as in \cite{Benjamin,Bona} we obtain
\beq\delta_h^2M\geq\frac{1}{4}\int_{-\infty}^{+\infty}\left({h^\prime}^2+\mathcal{C}h^2\right)dx-\frac{2}{5}\mathcal{C}^{\frac{1}{4}}
{\|\left(h,\xi\right)\|}^3_{H_1} \eeq and
\[\|h\|_{H_1}\geq d_{II}\left(u,h\right).\]

We will now obtain a lower bound for $\delta_\xi^2M.$

We consider the operator $H=-\frac{d^2}{dx^2}-\frac{1}{2}\phi$
with domain in the Hilbert space $L_2( \mathbb{R}).$ It is a
symmetric essentially self-adjoint operator. We denote with the
same letter $H$ its self-adjoint extension. It has two eigenvalues
$\lambda_1=-\mathcal{C}$ and $\lambda_2=-\frac{\mathcal{C}}{4}$
and a continuous spectrum $[0,\infty).$ The eigenfunctions are
proportional to $\frac{1}{\cosh^2z}$ and $\frac{\sinh
z}{\cosh^2z}$ respectively, where
$z=\frac{1}{2}{\mathcal{C}}^{\frac{1}{2}}$. The spectral theorem
for self-adjoint operators ensures the existence of an unitary
transformation from the domain $D(H)$ in the Hilbert space $
\mathcal{H}$ to $L_2( \mathbb{R},d\rho).$ In the case in which $
\mathcal{H}=L_2( \mathbb{R})$ this unitary transformation may be
realized in terms of the eigenfunctions $\psi_1,\psi_2\in L_2(
\mathbb{R})$ and the hypergeometric functions $\psi_\lambda(x)$
which satisfy point to point
\[H\psi_\lambda=\lambda \psi_\lambda\] for $\lambda>0$, but do not
belong to $L_2( \mathbb{R})$. Under the unitary map $f(x)\in L_2(
\mathbb{R}) $ can be expressed
\[f(x)=F_1\psi_1(x)+F_2\psi_2(x)+\int_0^\infty
F(\lambda)\psi_\lambda(x) d\rho(\lambda)\] where $F(\lambda)$
belongs to $L_2( \mathbb{R}^+,d\rho)$. The action of $H$ in $
\mathcal{H}$ corresponds to the multiplication by $\lambda$ in
$L_2( \mathbb{R}^+,d\rho)$:
\[Hf=\lambda_1F_1\psi_1+\lambda_2F_2\psi_2+\int_0^\infty \lambda F(\lambda)\psi_\lambda
d\rho(\lambda).\] We notice that for $f\in D(H)$ the third term
on the right hand member belongs to $L_2( \mathbb{R})$.
$\psi_\lambda$ are normalized in order to have
${\|f\|}_{\mathcal{H}}^2=F_1^2+F_2^2+\int_0^\infty
F^2(\lambda)d\rho(\lambda).$ $\psi_1,\psi_2,\psi_\lambda$ are
pairwise orthogonal with the internal product in $L_2(
\mathbb{R}).$

If we denote $g(x)=\int_0^\infty
F(\lambda)\psi_\lambda(x)d\rho(\lambda)$, $g\in L_2(
\mathbb{R})$, it follows
\[\int_{-\infty}^{+\infty}Hgdx=\int_{-\infty}^{+\infty}\left[\int_0^{\infty}\lambda
F(\lambda)\psi_\lambda d\rho(\lambda) \right]dx,\]

its left hand member is zero, hence

\beq\int_{-\infty}^{+\infty}\left[\int_0^{\infty}\lambda
F(\lambda)\psi_\lambda d\rho(\lambda) \right]dx=0\eeq for any
$F(\lambda)$ on the space $L_2( \mathbb{R}^+,d\rho)$.

We consider first $F(\lambda)$ to have support in the complement
of a neighborhood of $\lambda=0$. We may then consider $\lambda
F(\lambda)=\chi(\lambda)$ where $\chi(\lambda)$ is the
characteristic function with value one in the interval $(
\hat{\lambda}-\epsilon,\hat{\lambda}+\epsilon)$ and $0$
otherwise. We have
\beq\int_{-\infty}^\infty\int_{\hat{\lambda}-\epsilon}^{\hat{\lambda}+\epsilon}\psi_\lambda
d\rho(\lambda)dx=0 \eeq for any $ \hat{\lambda}$ in the support of
$F( \lambda)$.

It then follows, by decomposing $F(\lambda)$, that

\beq \int_{-\infty}^{+\infty}\left[\int_0^{\infty}
F(\lambda)\psi_\lambda d\rho(\lambda) \right]dx=0.\eeq Since the
$F(\lambda)$ considered is dense in $L_2( \mathbb{R}^+,d\rho)$ we
conclude that (31) is valid for any $F(\lambda)\in L_2(
\mathbb{R}^+,d\rho).$

We may always decompose $\xi\in L_2(\mathbb{ \mathbb{R}})$ as
\[\xi=F_1\psi_1+F_2\psi_2+\int_0^\infty F(\lambda)\psi_\lambda d\rho(\lambda)\]
and use (18) together with (31) to obtain

\beq F_1=0.         \eeq

We now consider the Rayleigh quotient of any $\xi$ in the domain
of $H$:

\[R(\xi)\equiv \frac{\left<-\xi,H\xi\right>}{\left<-\xi,\xi\right>}.\]

$\left<,\right>$ denotes the internal product in the $L_2(
\mathbb{R})$ space.

The eigenvalues satisfy (using the min-max theorem)

\[\lambda_1=\inf\left\{R(\xi):\xi\ \in \mathrm{\:domain\:} H\right\}\]

\[\lambda_2=\inf\left\{R(\xi):\xi\ \perp \mathrm{\:span\:}(\psi_1)  \right\}.\]

We thus get
\[\sum_i\left<\varphi_i,H\varphi_i\right>\geq \lambda_2\sum_i\left<\varphi_i,\varphi_i\right>.\]

From (27), and using $\lambda_2=-\frac{\mathcal{C}}{4}$, we obtain
\[\delta_\xi^2M\geq\frac{3}{4}\mathcal{C}\sum_i\left<\varphi_i,\varphi_i\right>\]
and introducing a parameter $0\leq\beta\leq1$

\begin{eqnarray*}\delta_\xi^2M &=& \left(1-\beta\right)\delta_\varphi^2M+\beta\delta_\varphi^2M\geq\left(1-\beta\right)\sum_{i=0}^\infty
\int_{-\infty}^{+\infty}\left({\varphi_i^\prime}^2+\mathcal{C}\varphi_i^2\right)dx+
\\ &+&
\left(1-\beta\right)\sum_{i=0}^\infty\int_{-\infty}^{+\infty}\left(-\frac{1}{2}\phi\right)\varphi_i^2dx+\frac{3}{4}\beta\mathcal{C}\sum_{i=0}^\infty
\varphi_i^2.
\end{eqnarray*}

Consequently, using $\phi\leq3\mathcal{C}$, we have
\[\delta_\xi^2M\geq\left(1-\beta\right)\min\left(1,\mathcal{C}\right){\|\xi\|}_{H_1}^2\]
for any $\beta$ satisfying $\frac{2}{3}<\beta<1$. In particular
for $\beta=\frac{3}{4}$

\beq\delta_\xi^2M\geq \frac{1}{4}
\min\left(1,\mathcal{C}\right){\|\xi\|}_{H_1}^2.\eeq

From (28) and (33) we get \beq\delta_h^2M+\delta_\xi^2M\geq
\frac{1}{4}\min\left(1,\mathcal{C}\right){\|\left(h,\xi\right)\|}_{H_1}^2-\frac{2}{5}\mathcal{C}^{\frac{1}{4}}
{\|\left(h,\xi\right)\|}_{H_1}^2\eeq

and from (21)

\[\delta^3M\geq-\frac{1}{2\sqrt{2}}{\|\left(h,\xi\right)\|}_{H_1}^2.\]

Finally,

\beq\Delta
M\geq\frac{1}{4}l{\|\left(h,\xi\right)\|}_{H_1}^2-b{\|\left(h,\xi\right)\|}_{H_1}^3\eeq
where \[l=\min\left(1,\mathcal{C}\right),
b=\frac{1}{2\sqrt{2}}+\frac{2}{5}\mathcal{C}^{\frac{1}{4}}.\]

We now use an argument in \cite{Benjamin,Bona} to show, for
$\delta$ small enough, that
\[\Delta M\geq\frac{1}{6}l{\|\left(h,\xi\right)\|}_{H_1}^2.\]

Since \[\|\left(h,\xi\right))\|_{H_1}\geq d_{II}\left[\left( u,\xi
\right),\left(\phi,0\right)\right]\] we then obtain
\[\Delta M\geq\frac{1}{6}l{\left\{ d_{II}\left[(u,\xi),(\phi,0)\right] \right\}}^2 \]
for $\delta$ small enough.

The stability proof has then been completed.

We may now relax the assumptions (17) and (18). The assumption
(17) may be removed by an application of the triangle inequality
as is \cite{Benjamin}. While (18) may be relaxed by considering
\[\tilde{\xi}=\xi-F_1\psi_1,\] which satisfies $\int_{-\infty}^\infty \tilde{\xi}dx=0$.

 Using again the triangle inequality for the distance $d_{II}$ we
 get
\[d_{II}\left[\left( u,\xi
\right),\left(\phi,0\right)\right]\leq d_{II}\left[\left(
u,\tilde{\xi}
\right),\left(\phi,0\right)\right]+d_{II}\left[\left( u,\xi
\right),\left(u,\tilde{\xi}\right)\right]\] where
$d_{II}\left[\left( u,\xi
\right),\left(u,\tilde{\xi}\right)\right]=\|F_1\psi_1\|_{H_1}$ is
conserved by $ \hat{H}_{\frac{1}{2}}$ and bounded by
$\|(u,\xi)\|_{H_1}$ at $t=0$. The solitonic solution $(\phi,0)$
is then stable in the sense (13), (14).

\section{Conclusions}
 Following \cite{Adrian} we considered the breaking of the
 supersymmetry in the $N=1$ Super KdV system and analize a solitonic model in terms of
 a Clifford algebra valued field. We showed the stability in the
 Lyapunov extended sense of the ground state and the one-soliton
 solution of the integrable model. The approach introduced in \cite{Benjamin} to prove the stability of the solitonic
 solutions of KdV equation and extended here for Clifford valued fields may also be used in the stability
 analysis of supersymmetric solitons in the bosonization scheme in \cite{Gao1,Gao2}.

 Independently of the original
 motivation, the breaking of supersymmetry, the system we analized in this paper is
 interesting because it contains the same soliton solutions as the
 KdV equation but is more realistic in the sense that the symmetry
 associated to the infinite number of conserved charges of KdV is
 broken.

$\bigskip$

\textbf{Acknowledgments}

A. R. and A. S. are partially supported by Project Fondecyt
1121103, Chile.

\end{document}